\documentclass[epj]{svjour}
\usepackage{graphicx}
\begin{document}
\title{Effect of structural distortions on the magnetism of doped spin-Peierls CuGeO$_3$}
\author{V. Simonet\inst{1} \and B. Grenier\inst{2,3} \and F. Villain\inst{4} \and A.-M. Flank\inst{5} \and G. Dhalenne\inst{6} \and A. Revcolevschi\inst{6} \and J.-P.
Renard\inst{7}}
%
%
\institute{Laboratoire Louis N\'eel, UPR 5051, B.P. 166, F-38042
Grenoble Cedex 9, France \and CEA-Grenoble, DRFMC / SPSMS / MDN,
F-38054 Grenoble Cedex 9, France \and Universit\'e Joseph Fourier,
B.P. 53, F-38041 Grenoble Cedex 9, France \and Laboratoire de
Chimie Inorganique et Mat\'eriaux Mol\'eculaires, UMR 7071, UPMC,
4 place Jussieu, F-75252 Paris Cedex 05, France \and SOLEIL,
L'Orme des Merisiers, Saint-Aubin - BP 48, F-91192 Gif-sur-Yvette
Cedex, France \and Laboratoire de Physico-Chimie de l'Etat Solide,
UMR 8648, Universit\'e Paris-Sud, B\^at. 414, F-91405 Orsay Cedex,
France \and Institut d'\'electronique Fondamentale, UMR 8622,
B\^at. 220, Universit\'e Paris-Sud, F-91405 Orsay Cedex, France}
\date{Received: date / Revised version: date}
%
\abstract{ The chemical selectivity and great sensitivity of the
Extended X-ray Absorption Spectroscopy technique allowed the
determination, in the paramagnetic phase, of the structural
distortions induced by doping in the spin-Peierls CuGeO$_3$
compound. The distorted environments were analyzed as a function
of concentration, magnetic nature of impurity and the substitution
site (Ni, Mn and Zn impurities on the Cu site, Si impurity on the
Ge site). This has led to estimate the variation of the angles and
pair distances, and hence to evaluate the magnetic coupling along
the Cu chains in the vicinity of the impurities. The
antiferromagnetic interaction between Cu first neighbors in the
pure sample is found to be weakened around Ni, almost cancelled in
the case of Mn doping, and even to change sign, producing a
ferromagnetic coupling for Si doping. More generally, the
structural distortions on a local scale are shown to be key
parameters for the understanding of the magnetic properties of
doped spin-Peierls compounds.
} 
\maketitle
\section{Introduction}
\label{intro}

\begin{figure}[t]
\[\includegraphics[scale=0.5]{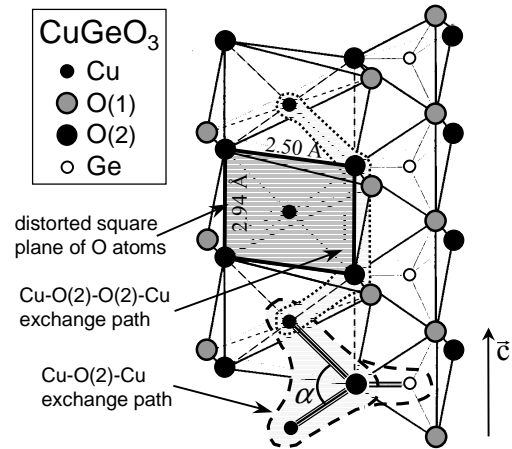}\]
\caption{Schematic structure of CuGeO$_3$} \label{Struc1}
\end{figure}

The first inorganic spin-Peierls (SP) compound,
CuGeO$_3$,\cite{hase} has already been extensively studied for its
original magnetic properties. The possibility of substituting the
cationic sites, Cu or Ge, by different magnetic or non-magnetic
elements, has brought new information and rose new questions.

The structure of CuGeO$_3$ can be described as a stacking of
CuO$_6$ octahedra and GeO$_4$ tetrahedra along the $c$ direction
of the structure. Each O belongs to both a Ge tetrahedron and a Cu
octahedron. The stacking of distorted octahedra (elongated along
the direction of the apical O(1) and with a planar distorted
square of basal O(2) in the perpendicular direction) results in
chains of magnetic Cu$^{2+}$ ions located at the center of edge
sharing squares of O(2) running along the $c$ axis (Fig.\
\ref{Struc1}). At room temperature, this compound then consists of
uniform chains of spins $S$=1/2, interacting by superexchange
through O(2) atoms leading to an antiferromagnetic (AF)
Heisenberg-type coupling (Fig.\ \ref{Struc1}). A magnetoelastic
instability produces a dimerization of the chains below a critical
temperature $T_{SP}$ of 14.25 K. This leads to a singlet
(non-magnetic) ground state separated from the first excited
triplet state ($S=1$) by an energy gap of 2 meV. Below $T_{SP}$,
the magnetic susceptibility drops to zero with a temperature
activated behavior, and diffraction experiments reveal new
superlattice peaks \cite{pouget}.

\begin{figure}[t]
\[\includegraphics[scale=0.5]{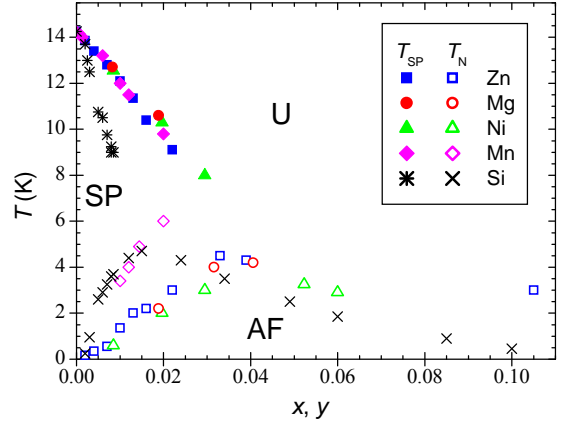}\]
\caption{(Temperature - concentration) magnetic phase diagram for
CuGe$_{1-x}$Si$_x$O$_3$ and Cu$_{1-y}M_y$GeO$_3$, with $M$=Zn, Mg,
Ni, and Mn.}\label{Diag}
\end{figure}

Doping, either with Zn, Ni, Mg or Mn on the Cu site
\cite{subCu,Coad,grenier1} or with Si on the Ge site
\cite{renard,grenierSi}, results into a decrease of $T_{SP}$ and,
at a lower temperature, the N\'eel temperature $T_N$, into a
transition to an AF long-range ordering (LRO). Some of these
impurities are magnetic, like Ni with spin 1 and Mn with spin 5/2
\cite{grenier3}, while the others, Zn, Mg and Si, are non
magnetic. They also have very different ionic radii whose
consequence on the magnetic properties will be discussed in Sec.
IV and V. A systematic study of the magnetic properties of single
crystals with various doping levels (CuGe$_{1-x}$Si$_x$O$_3$ with
$0 < x \leq 0.08$, and Cu$_{1-y}M_y$GeO$_3$ with $M$ = Zn, Mg, Ni
and $0 < y \leq 0.1$), all synthesized and characterized in the
same way, has allowed establishing a quasi-universal phase diagram
for the spin-Peierls transition \cite{grenier1}. The variation of
$T_{SP}$ versus concentration was found identical for all
impurities on the Cu site, including Mn, which has been studied
since Ref. \cite{grenier1} (for $0 < y \leq 0.02$), and for Si
impurities applying a scaling factor $y \approx 3x$. This is
illustrated in Fig. 2 of Ref. \cite{grenier1} and in the
alternative representation of the present Fig. \ref{Diag} where no
scaling factor $y \approx 3x$ is applied in view of the discussion
of Sec. V.B. Surprisingly, doping outside the spin chains is thus
about three times more efficient in destroying the spin-Peierls
phase than doping within the chains \footnote{Note that, while for
Ni, Zn and Mg, the results of Grenier {\it et al.} \cite{grenier1}
are in rather good agreement with the work of other groups on
crystals of different origin, a large spread of results exists
concerning Si-doping. The slope of $T_{SP}(x)$ from the works of
Katano {\it et al.} \cite{katano} and Hiroi {\it et al.}
\cite{hiroi} is smaller than that found by Grenier {\it et al. }
while it is similar in the work of Wang {\it et al.}\cite{wang},
leading to the scaling factors $y=x$, $y=1.5~x$ and $y\sim2-3~x$
respectively. However, we are more confident in the results of
Grenier {\it et al.} concerning the Si-doping data since the
weaker effect measured by other groups suggests that not all Si
ions enter the CuGeO$_3$ structure and/or part of them segregates
(as was already observed in highly doped samples). This leads to
overestimate $x$ and thus results in an apparent smaller scaling
factor between $y$ and $x$.}. In contrast, $T_N$ varies with the
type of impurity: it is higher for Mn and Si-doping and slightly
lower for Ni-doping, in comparison with the value, found
identical, for the non-magnetic impurities Mg and Zn. A difference
between Ge site and Cu site doping is observed concerning the
$T_N(y,x)$ transition line which is continuous in the first case
and presents a kink between a dimerized AF phase at low doping and
a uniform AF phase at larger doping in the second case
\cite{masuda}. This behaviour is characteristic of a second-order
and a first-order phase transition, respectively. Another
particularity concerns the easy axis of magnetization in the AF
phase: it is the $a-$axis for Ni doping and the $c-$axis for all
other substitutions, which can be explained by the single-ion
anisotropy of Ni$^{2+}$ in a distorted octahedral environment
\cite{grenier2}.

These results must be considered in the light of the different
couplings along the Cu chains. The magnetic properties are mainly
driven by two kinds of AF couplings: A superexchange coupling
Cu-O(2)-Cu between first Cu neighbors with an AF exchange constant
$J_1 = 120-180$ K and a frustrating AF coupling between second
neighbors $J_2=0.24-0.36~J_1$ via the Cu-O(2)-O(2)-Cu path
\cite{castilla,nishi} (Fig.\ \ref{Struc1}). The inter-chain
couplings are one order and two orders of magnitude smaller along
the $b$ and $a$ axes, respectively \cite{nishi}. These inter-chain
couplings induce a coherence between the phases of the distortion
waves on different chains. In CuGeO$_3$, the dimerization pattern
is out of phase for adjacent chains in the $b$ and $a$ directions.
The strength and nature of the magnetic interactions are strongly
dependent on the Cu-O(2) and Ge-O(2) distances and even more
drastically on the Cu-O(2)-Cu superexchange angle (denoted
$\alpha$). The Goodenough-Kanamori-Anderson rules
\cite{goodenough} state that the exchange is ferromagnetic (FM)
for an angle of 90$^{\circ}$. The deviation of the Cu-O(2)-Cu
angle from this value ($\alpha$=99$^{\circ}$ in CuGeO$_3$) will
drive the exchange towards antiferromagnetism \cite{braden}.
However, Geertsma and Khomskii pointed out that calculations
taking into account this sole effect still yield a slightly
ferromagnetic exchange coupling \cite{khomskii1}: The
hybridization of Ge orbitals with the 2$p_y$ ones of the O(2),
which destroys the equivalence of 2$p_x$ and 2$p_y$ O(2) orbitals,
must indeed be taken into account to obtain, for $J_1$, an AF
exchange with a reasonable value of 135~K.

The aim of the present work is to understand the effect of doping
the Cu chains, with magnetic and non-magnetic impurities, and out
of the chains in the case of Si-doping, by measuring the induced
structural distortions of doped CuGeO$_3$. These distortions
should strongly influence the magnetic properties, in particular
via the variation of the superexchange angle $\alpha$ and thus of
the related magnetic coupling $J_1$. The EXAFS (Extended X-ray
Absorption Fine Structure) spectroscopy technique was chosen for
this purpose as a local order probe, with chemical sensitivity,
and for its possibility to measure very diluted systems. For the
first time with this technique, the influence of Ni, Mn, Zn and
Si-doping was investigated in the paramagnetic phase ($T >
T_{SP}$) but not that of Mg-doping because of the too low energy
absorption of the Mg atom.

The next section is devoted to the experimental details and to a
brief recall of the EXAFS analysis methodology. The EXAFS results
are reported in section III for Cu site and Ge site doping. The
structural distortions, as deduced from the EXAFS refined
parameters, are analyzed in section IV. Their influence is
discussed in section V, in relation with the magnetic properties,
and a conclusion is given in Sec. VI.

\section{Experiment}
\label{exp}
\begin{figure}
\[\includegraphics[scale=0.45]{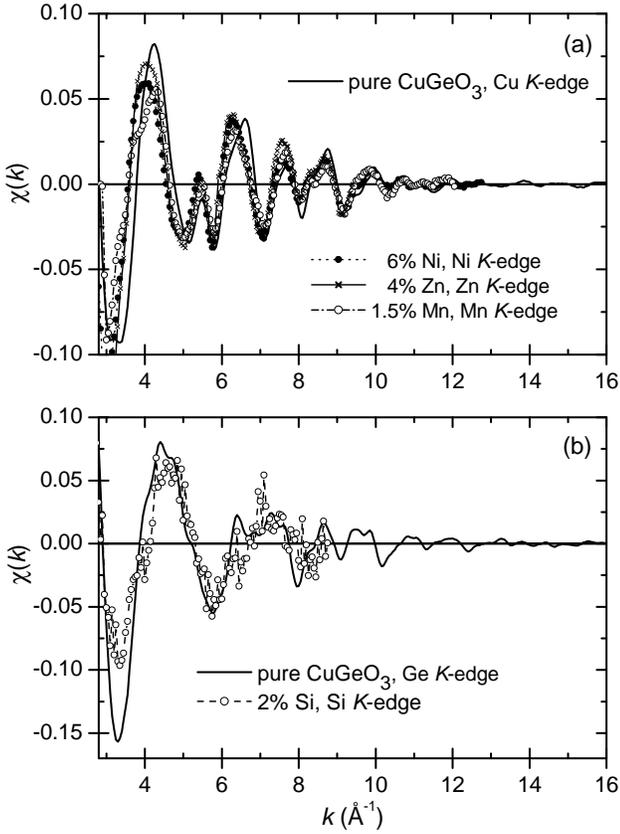}\]
\caption{Comparison of the measured $\chi(k)$ functions (a) for
pure CuGeO$_3$ at the Cu $K$-edge, and for 6\% Ni, 1.5\% Mn and
4\% Zn doped samples, at the impurity $K$-edges and (b) for pure
CuGeO$_3$ at the Ge $K$-edge and for 2\% Si-doped sample at the Si
$K$-edge.} \label{exafs1}
\end{figure}

The EXAFS measurements at the Cu and Ge absorption $K$-edges, in
transmission detection mode, and at the Ni, Mn and Zn absorption
$K$-edges, in fluorescence detection mode, were carried out at
LURE (Orsay) on the D42 and D44 beamlines of the DCI storage ring.
The experiments were performed at 78 K, in the paramagnetic phase,
above the spin-Peierls transition. The incident energy was varied
by 2 eV steps using a Si(111) single crystal monochromator. The
EXAFS oscillations were obtained in transmission detection mode by
measuring the X-ray beam intensities with ionization chambers,
before ($I_0$) and after ($I_1$) the sample, in an energy range
close to the absorption $K$-edge of Cu ($\lambda_K^{Cu}$=8979 eV)
and Ge ($\lambda_K^{Ge}$=11103 eV). The samples studied in the
transmission detection mode were pellets of homogeneous thickness
made of a mixture of about 15 mg of single crystal ground into
very fine powder and 35 mg of cellulose. In the fluorescence
detection mode, the powder obtained from ground doped crystals was
spread on an adhesive tape. The EXAFS spectra were measured in the
fluorescence detection mode with a multielements detector around
the impurities $K$-edges in the 2\% and the 6\% Ni-doped samples
(absorption $K$-edge $\lambda_K^{Ni} = 8333$ eV), 1.5\% Mn-doped
sample ($\lambda_K^{Mn} = 6539$ eV), and the 4\% Zn-doped sample
($\lambda_K^{Zn} = 9659$ eV).

The Si $K$-edge EXAFS experiment was performed on the SA32
beamline of SUPERACO using a 2 InSb crystals monochromator. A thin
slice of 2\% Si-doped single crystal (absorption $K$-edge
$\lambda_K^{Si} =$ 1839 eV) was cleaved and stuck with Ag glue on
an indium covered sample-holder. The absorption signal was
recorded with a monoelement detector in the fluorescence detection
mode.

The EXAFS oscillations were analyzed using the FEFF package
\cite{feff}. The EXAFS oscillations, $\chi(k)$, were extracted
from the experimentally measured absorption coefficient $\mu(E)$
using the AUTOBK program with $k$, the wavenumber of ejected
photoelectron, equal to:
\begin{equation}
\label{equation1} k = \sqrt{2m(E-E_0)/\hbar^2}
\end{equation}
\noindent where $E_0$ is the absorption energy for the $K$-edge
and $m$ the electron mass.

For an absorbing central atom surrounded by $N$ identical atoms,
when considering only single scattering paths, EXAFS oscillations
created by the backscattering of the photoelectron from the
neighboring atoms are given by:
 \begin{eqnarray}
 \label{equation2}
 \chi(k) = - S_0^2 \frac{N}{kr^2}~\vert F\left(k,\pi\right)\vert ~e^{-2k^2\sigma^2}e^{-2r/\lambda(k)} \nonumber \\
 \sin\left[2kr+2\delta(k)+\Phi(k)\right]
 \end{eqnarray}
\noindent where $r$ is the mean pair distance, $\sigma^2$ is the
associated Debye-Waller factor which takes into account the
structural and dynamical disorders, and $S_0^2$ is the amplitude
reduction factor accounting for many-body effects within the
absorbing central atom. The other parameters in Eq.
(\ref{equation2}) are the phase $\Phi(k)$ and the modulus of the
complex function $F(k, \pi)$, which both depend on the nature of
the backscattering atom, the mean free path $\lambda(k)$ and the
absorbing atom phase shift $\delta(k)$. They are derived from the
theoretical ab-initio multiple scattering calculations by the FEFF
program using crystallographic data. In order to describe a real
environment, the surrounding atoms are gathered into atomic shells
(atoms at approximately the same distance from the absorber),
themselves subdivided into several subshells, for each kind of
chemically and crystallographically equivalent neighbors.
\begin{figure}
\[\includegraphics[scale=0.5]{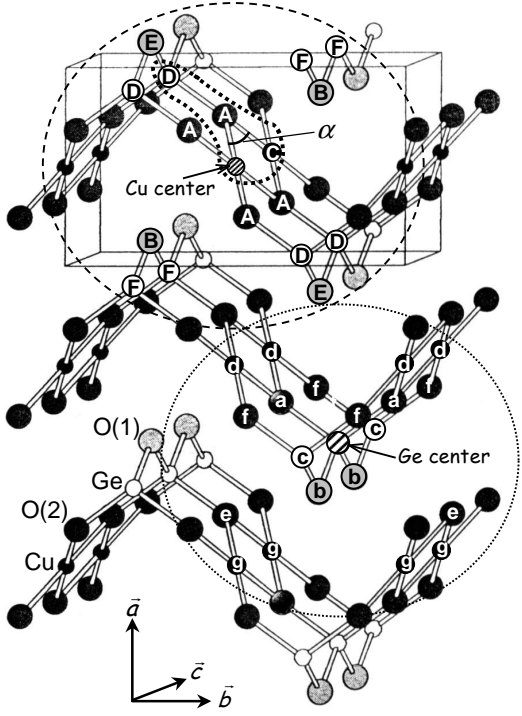}\]
\caption{Structure of CuGeO$_3$. The Cu and Ge environments are
materialized: starting from the dashed Cu and Ge absorbers, the
atoms labelled "A","B", ... "F" (for Cu) and "a", "b", ... "g"
(for Ge) are their neighbors going from the closest to the
furthest ones that were accessible in the EXAFS experiment.}
\label{Struc2}
\end{figure}

Representative measured EXAFS oscillations $\chi(k)$ are shown in
Fig.\ \ref{exafs1}. Note that the spectra of the doped samples,
recorded at the impurity $K$-edge, are more noisy than those
recorded at the Cu and Ge $K$-edges because of the small amount of
absorbers and of the fluorescence detection mode. In consequence,
the EXAFS analysis could not be performed up to such large $k$
values at the impurity $K$-edges compared to the Cu and Ge ones.
For the Si $K$-edge spectrum, an additional source of noise was
the low fluorescence yield and the small photon flux at this low
energy. The complex Fourier transforms (denoted FT) of the $k^3$
weighted $\chi(k)$ functions were then calculated using a Hanning
window. The amplitude of this FT can be associated with a
pseudo-radial atomic distribution around the absorber in the
$r$-space. Lastly, the EXAFS oscillations of the atomic shells of
interest have been isolated by a subsequent inverse
Fourier-transform (denoted IFT) performed in a restricted
appropriate $r$-range and leading to a Fourier filtered
$k^3\chi(k)$ function. The FT and IFT were simulated with the
FEFFIT program by summing the contributions given in Eq.
(\ref{equation2}) for all subshells obtained from the structural
model. The multiple scattering paths, whose contribution was
checked to be weak, were not included in the calculation. In the
fit, the pair distances, the Debye-Waller factors and the energy
shift $\Delta E_0$ were allowed to vary \footnote{$\Delta E_0$
accounts for inaccuracies in the FEFF determination of the
absorption threshold energy $E_0$. Different values of $\Delta
E_0$ were refined for the O, Cu and Ge neighbors. These extra
shifts are meant to correct errors introduced by FEFF in the
calculation of the scattering phase shifts.}. For the analysis of
the EXAFS oscillations recorded at the impurity absorption edges,
the FEFF files for each scattering path were recalculated with the
impurity replacing Cu or Ge absorber within the same structure.

\section{Results}

\subsection{Pure sample}
\begin{figure}
\[\includegraphics[scale=0.5]{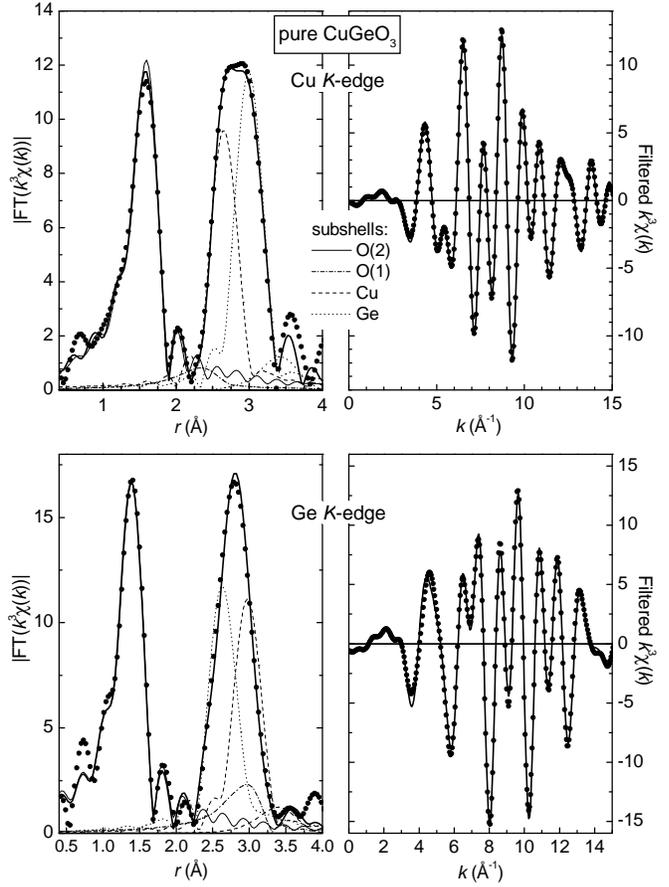}\]
\caption{EXAFS results for the pure CuGeO$_3$ sample. Amplitude of
the FT of $k^3\chi(k)$ (left) and Fourier filtered $k^3\chi(k)$
(right) at the Cu (top) and Ge (bottom) $K$-edges. The measured
spectra are represented by full circles, the fit by a thick solid
line. The contributions of each kind of neighbors are also shown
on the FT plots (left).} \label{exafs2}
\end{figure}
The structure of CuGeO$_3$ in the paramagnetic state was initially
refined by V\"ollenkle {\it et al.} \cite{vollenkle} and later by
Braden {\it et al.} \cite{braden,braden2} (space group $Pbmm$,
$a=4.796$~\AA, $b=8.466$~\AA, $c=2.940$~\AA~at 295 K), from
Rietvelt refinement of X-ray and neutron powder diffractograms
\footnote{The structure was revised by M. Hidaka {\it et al.}, J.
Phys.: Condens. Matter {\bf 9}, 809 (1997). They found a different
space group $P$2$_1$2$_1$2$_1$ and a larger unit cell $2a\times b
\times 4c$ than Braden {\it et al.} \cite{braden}, who nonetheless
reconfirmed the initial structure one year later \cite{braden2}.}.
The crystallographic data of Braden {\it et al.} \cite{braden}
were used as input of the FEFF program in our EXAFS analysis.
These structural parameters yield, for {\it the atomic
distribution around Cu atoms}, a first shell with the 4 square
planar O(2) at 1.942 \AA~ and a second shell including the 2
apical O(1) at 2.78 \AA, the 2 chain neighbors Cu at 2.95 \AA~ and
4 Ge at 3.30 \AA~ (top of Fig.\ \ref{Struc2}). The {\it atomic
distribution around Ge atoms} is made of a first shell with the 4
O of the tetrahedron (2 O(2) at 1.737 \AA~ and 2 O(1) at 1.778
\AA) and a second shell including 2 Ge at 2.946 \AA~ and 4 Cu at
3.30 \AA~ (cf. bottom of Fig.\ \ref{Struc2}). This shell
description is summarized in Table I.

\begin{table*}
\caption{Structural parameters of the atomic radial distribution
up to 3.95 \AA centered on Cu and Ge atoms, from the
crystallographic data of Braden et al. \cite{braden} and from the
EXAFS simulations of the pure CuGeO$_3$ sample at both Cu and Ge
$K$-edges. The atom labels are those of Fig. \ref{Struc2}.
Crystallographic data of Braden {\it et al.} \cite{braden} imply
two slightly different Ge-O distances for the two pairs of oxygens
within the tetrahedron. The parameters reported with no error bar
were held fixed during the fit. The error bars only account for
the statistical uncertainties and are therefore probably
underestimated. The pair distances $r$ and Debye-Waller factors
$\sigma^2$ are in \AA~and \AA$^2$ respectively. The quality of the
EXAFS fit is given by the $R$-factor: $Rf=$0.002 at the Cu
$K$-edge and $Rf=$0.008 at the Ge $K$-edge. The refined $S_0^2$
values are 0.927 at the Cu $K$-edge and 0.953 at the Ge $K$-edge.}
 \label{table1}
 \vspace{0.5cm}
 \begin{tabular}{ll|l|ll||ll|l|ll}
 \hline\hline
 \multicolumn{2}{c|}{\bf{Cu center}} & Ref. \cite{braden} & \multicolumn{2}{c||}{EXAFS} & \multicolumn{2}{c|}{\bf{Ge center}} & Ref. \cite{braden} & \multicolumn{2}{c}{EXAFS}\\
 \hline
  atom & label & $r$ & $r$ & $\sigma ^2$& atom & label & $r$ & $r$ & $\sigma ^2$ \\
 \hline
  4 O(2) & ~A & 1.9324  & 1.942(5)  & 0.0032(4) & 2 O(2) & ~a & 1.7322 & 1.737(3)  & 0.0022(3)\\
  2 O(1) & ~B & 2.7547  & 2.77(4)   & 0.010(6)  & 2 O(1) & ~b & 1.7729 & 1.778(3)  & 0.0022(3)\\
  2 Cu   & ~C & 2.9400  & 2.944(7)  & 0.0027(4) & 2 Ge   & ~c & 2.9400 & 2.941(5)  & 0.0022(3)\\
  4 Ge   & ~D & 3.2872  & 3.294(7)  & 0.0040(2) & 4 Cu   & ~d & 3.2872 & 3.294(7)  & 0.0040(3)\\
  2 O(1) & ~E & 3.6975  &           &           & 2 O(2) & ~e & 3.3015 & 3.301     & 0.006(4)\\
         & ~  &         &           &           & 4 O(2) & ~f & 3.4123 & 3.412     & 0.006(4)\\
  4 Ge   & ~F & 3.7717  & 3.74(3)   & 0.014(2)  & 4 Cu   & ~g & 3.7717 & 3.74(3)   & 0.014(2) \\
 \hline\hline
\end{tabular}
\end{table*}

The EXAFS fitting procedure was tested on the pure CuGeO$_3$
sample. The first two atomic shells of the radial distributions at
both Cu and Ge $K$-edges were simultaneously fitted. This implies
that the distance and the Debye-Waller factor corresponding to
each Ge-Cu paths were hold identical for the two sets of data
during the refinement. The $k$ range used for the FT and for the
fitting procedure was [3.5, 15 \AA$^{-1}$] while the $r$ range
used for the IFT was [1, 3.4 \AA] at both edges. The fits are in
good agreement with the experimental spectra, as shown in Fig.\
\ref{exafs2}, where the contribution of each kind of neighbors is
represented. Note that, at the Cu $K$-edge, the signal from the
two apical oxygens O(1) is already very weak and the further O(1)
neighbors were not considered because yielding a negligibly small
contribution. However, the presence of the next 4 Ge atoms at
about 3.786 \AA~was found necessary to fit correctly the second
shell. The resulting structural parameters obtained at both edges
are listed in Table I and the comparison with their initial values
shows the consistency of the fit. After this first EXAFS analysis,
the same procedure was applied to the doped samples.

\subsection{Substitution on the Cu site}

\begin{figure}
\[\includegraphics[scale=0.5,bb=30 70 562 780]{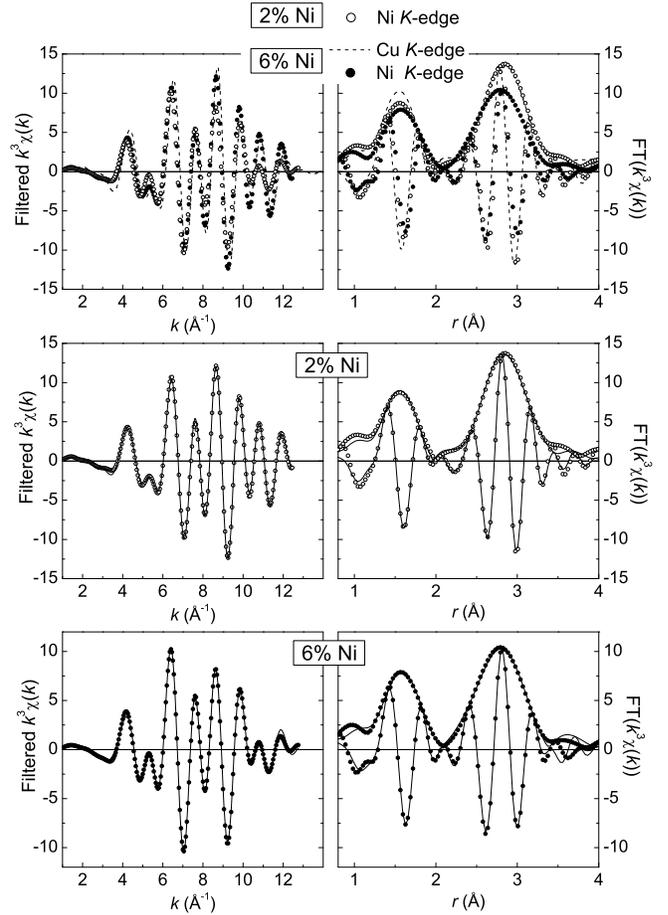}\]
\caption{EXAFS results for the 2\% and 6\% Ni-doped CuGeO$_3$.
Amplitude and imaginary part of the FT of $k^3\chi(k)$ (right) and
Fourier filtered $k^3\chi(k)$ (left). Comparison of the measured
spectra at the Ni $K$-edge for both doping levels and at the Cu
$K$-edge for 6\% Ni sample (top). Comparison of the measured
spectra (circles) and of the fits (solid lines) at the Ni K-edge
for 2\% doping (middle) and 6\% doping (bottom).} \label{exafs3}
\end{figure}

The Ni and Mn $K$-edges EXAFS oscillations were analyzed up to
$k=12$ \AA$^{-1}$ (see Fig.\ \ref{exafs1}). In the case of Zn, an
additional difficulty arose from the mixing of Cu $K$-edge and Zn
$K$-edge EXAFS oscillations due to the closeness of the absorption
energies of both elements: 8979 eV for Cu and 9659 eV for
Zn.\footnote{To correct the Zn $K$-edge spectrum, the Cu $K$-edge
oscillations $\chi(E)$ of the pure sample, recorded up to 9998 eV,
were converted into $\chi(k)$ using Eq.(\ref{equation1}) with the
absorption energy $E_0$ of Zn. The Cu EXAFS contribution, whose
amplitude at the Zn edge was obtained from the ratio of both
absorption jumps recorded in a unique scan in the doped sample,
was then subtracted from the Zn spectrum.}. Therefore, the Zn
$K$-edge spectrum, after correction from the contamination by the
Cu $K$-edge oscillations, could be analyzed in a more limited $k$
range, up to 9 \AA$^{-1}$ (cf. Fig.\ \ref{exafs1}).

First, the structural results derived at the Cu $K$-edge in the
doped samples were found identical to those of the pure sample,
even for doping levels as high as 6\% (Ni). The distortions
induced by the impurities were then deduced from the comparison of
the structural parameters obtained at the impurity and at the Cu
$K$-edge in the same sample and using the same $k$ and $r$-ranges
of analysis for the EXAFS oscillations. Note that we did not try
to take into account impurity neighbors in our structural model
(although there can be some for high doping levels) because it is
most probably impossible to distinguish them from Cu neighbors in
the EXAFS analysis.

\begin{figure}
\[\includegraphics[scale=0.5]{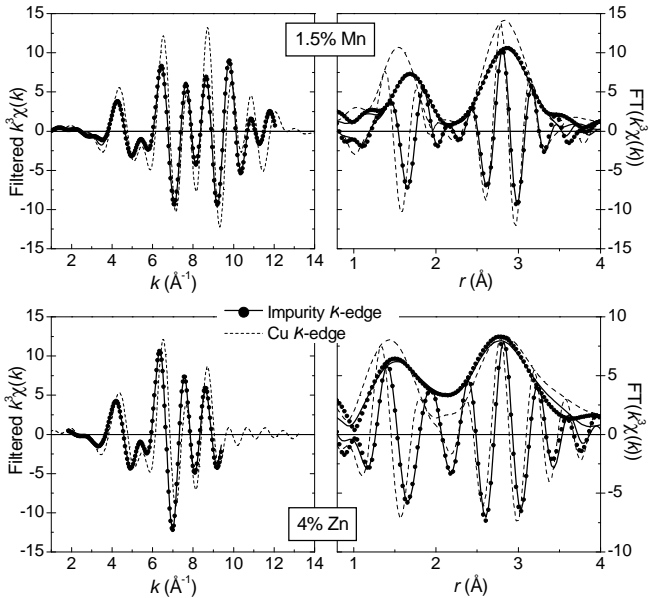}\]
\caption{EXAFS results for 1.5\% Mn (top) and 4\% Zn (bottom)
doped CuGeO$_3$ : amplitude and imaginary parts of the FT of
$k^3\chi(k)$ (right) and Fourier filtered $k^3\chi(k)$ (left) of
the measured data (full circles) and their fits (solid lines) at
the impurity $K$-edge. The Cu $K$-edge spectra measured on the
same sample are also shown (dashed lines), for
comparison.}\label{exafs4}
\end{figure}

The FT of $k^3\chi(k)$ and Fourier filtered $k^3\chi(k)$ are shown
in Fig.\ \ref{exafs3} for the 2\% and 6\% Ni-doped compounds and
in Fig.\ \ref{exafs4} for the 1.5\% Mn and 4\% Zn-doped compounds.
The spectrum recorded at the impurity $K$-edge in each compound is
compared to its fit and to the reference spectrum at the Cu
$K$-edge. The structural parameters extracted from the fits are
reported in Table II. The fits in $k$ and $r$ spaces are of rather
good quality. For all doping values, the amplitude of the FT, for
the first O shell, is smaller at the impurity $K$-edge than at the
Cu one, which is related to an increase of the Debye-Waller
factor, in agreement with a doping-induced structural disorder. An
increased Debye-Waller factor is also observed for the second
shell in the case of the 6\% Ni and 1.5\% Mn-doped compound. Note
that the 4\% Zn-doped compound shells are less well-defined
because of the smaller $k$ range of integration. For all
compounds, the first shell is shifted towards larger $r$ values.
The distance from the absorber to the O(2) first neighbors
increases more and more when replacing Cu by Ni ($\Delta r/r =$
1.4\% and 2.2\% for 2 and 6\% doping resp.), Zn ($\Delta r/r =$
4.5\%) and Mn ($\Delta r/r =$ 7.3\%). Interestingly, the
structural distortion varies also for a same impurity as a
function of its doping level, as shown by the increase of the
Debye-Waller factor and first Ni-O(2) distance from 2\% to 6\% Ni.
For Ni and Zn impurities, there is an ambiguity concerning the
position of the next nearest neighbors, the apical oxygens O(1).
They are found either closer or further away from the impurity
than from Cu (see parameters of the equally good EXAFS fits in
Table II). At the Mn $K$-edge, there is no such ambiguity since
the EXAFS spectra is 10 times better fitted when the O(1) atoms
are closer from the impurity than further away, as compared to the
Cu environment. Lastly, there is a clear variation of the
distances to the next Cu and Ge neighbors with respect to the Cu
environment (1.6\% and 1\% increase, respectively) in the case of
the Mn impurity only.

\subsection{Substitution on the Ge site}
\begin{figure}
\[\includegraphics[scale=0.5]{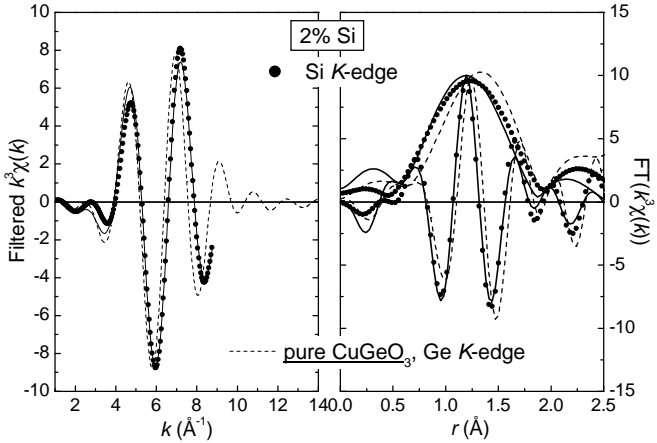}\]
\caption{EXAFS results for 2\% Si-doped CuGeO$_3$. Amplitude and
imaginary part of the FT of $k^3\chi(k)$ (right) and Fourier
filtered $k^3\chi(k)$ (left) at the Si $K$-edge of the doped
sample (full circles) compared to the fit (solid lines) and to the
results obtained at the Ge $K$-edge on the pure sample (dashed
lines).} \label{exafs5}
\end{figure}
\begin{table*}
\caption{Structural parameters obtained from the simulations of
the EXAFS spectra at the impurities $K$-edges of the Cu-site doped
CuGeO$_3$ samples: $R$-factors (quality of the fit), relative
distance variations $\Delta r/r$ and Debye-Waller factors
$\sigma^2$. The relative distance variations induced by impurities
are calculated with respect to the results obtained for the Cu
$K$-edge spectrum of the same sample analyzed using the same $k$
and $r$ ranges (see Table \ref{table1}). For Ni and Zn impurities,
two roughly equivalent solutions of the EXAFS fits are reported
with a different distance from the impurity to the apical oxygens
O(1), while only one solution is relevant for Mn. The results
given in bold characters correspond to the solutions validated by
the analysis presented in Sec. \ref{IV-A}. Note that the
parameters presented with no error bar were held fixed during the
fit and equal to the Cu $K$-edge refined value in the same sample.
The $S_0^2$ factors at all the impurities $K$-edges were fixed to
the value of 0.927 obtained at the Cu $K$-edge.} \label{table2}
\vspace{0.5 cm}
 \begin{tabular}{ll|ll|ll|ll|l}
 \hline\hline
       & & \multicolumn{2}{c|}{{\bf 2\% Ni}} & \multicolumn{2}{c|}{{\bf 6\% Ni}} & \multicolumn{2}{c|}{{\bf 4\% Zn}} & {\bf 1.5\% Mn}\\
  \hline
  $Rf$                    &        &~{\bf 0.0021}   &~0.0023   &~{\bf 0.0025}   &~0.0058   &~{\bf 0.0015}    &~0.0030   &~{\bf 0.0016}    \\
  \hline
                          & O(2)~~~&~~{\bf 1.4(7)}  &~1.3(6)   &~~{\bf 2.2(7)}  &~2.2(7)   &~~{\bf 4.5(1.8)} &~3.9(1.3) &~~{\bf 7.3(6)}   \\
                          & O(1)   &~-{\bf 17(4)}   &~4(4)     &~-{\bf 15(2)}   &~4(4)     &~-{\bf 15(3)}    &~9(2)     &~-{\bf 13(1)}    \\
  ${\Delta r \over r}$(\%)& Cu     &~~{\bf 0.6(6)}  &~0.7(7)   &~-{\bf 0.3(6)}  &~0.1(8)   &~-{\bf 0.9(8)}   &-0.7(1.0) &~~{\bf 1.6(6)}   \\
                          & Ge     &~~{\bf 0.0(4)}  &~0.15(70) &~~{\bf 0.1(5)}  &~0.4(7)   &~~{\bf 0.3(1.6)} &~1.3(1.2) &~~{\bf 1.0(6)}   \\
                          & Ge     &~~{\bf 0.0}     &~0.15     &~~{\bf 0.1}     &~0.4      &~~{\bf 0.3}      &~1.3      &~~{\bf 1.0}      \\
  \hline
                          & O(2)   &~{\bf 0.0038(4)}&~0.0039(4)&~{\bf 0.0042(4)}&~0.0042(5)&~{\bf 0.0035(11)}&~0.0044(6)&~{\bf 0.0045(3)} \\
                          & O(1)   &~{\bf 0.025(12)}&~0.017(13)&~{\bf 0.0119(4)}&~0.013    &~{\bf 0.017(13)} &~0.01     &~{\bf 0.0088(20)}\\
  $\sigma^2$ (\AA$^2$)    & Cu     &~{\bf 0.0034(8)}&~0.0032(9)&~{\bf 0.0038(6)}&~0.0035(8)&~{\bf 0.002}     &~0.002    &~{\bf 0.0047(10)}\\
                          & Ge     &~{\bf 0.0032(4)}&~0.0034(5)&~{\bf 0.0045(4)}&~0.0049(6)&~{\bf 0.0045}    &~0.0045   &~{\bf 0.0048(4)} \\
                          & Ge     &~{\bf 0.021(11)}&~0.025(19)&~{\bf 0.012(3)} &~0.014(6) &~{\bf 0.014}     &~0.014    &~{\bf 0.026(16)} \\
   \hline\hline
\end{tabular}
\end{table*}
The EXAFS oscillations, recorded at the Si $K$-edge in a 2\%
Si-doped sample, are limited to 8.7 \AA$^{-1}$ and allow getting
information only about the first O tetrahedral shell. Note that,
due to the small range of fitting, the Debye-Waller factor could
not be accurately refined and was fixed to 0.002 \AA$^2$, {\it
i.e.} to the value found in the pure sample at the Ge $K$-edge.
This may be one reason for the small discrepancies ($R$-factor
value of 0.011) observed between the fit and the experimental data
(Fig.\ \ref{exafs5}). The results, obtained using $S_0^2$=0.81 in
the fitting procedure, indicate a strong contraction of the Si-O
distance of about 8.4\% with respect to the Ge-O distance. In
addition, a more symmetric environment is found around the Si than
around the Ge atoms, that can be described by a small regular O
tetrahedron with distance from center to corner of 1.61(2)
\AA~(instead of 1.74 and 1.78 \AA~ for the two pairs of oxygen
forming the distorted tetrahedron surrounding the Ge atoms).

\section{Analysis}

\subsection{Comparison with ionic radii}
\label{IV-A}

The ionic radius for each chemical species, which depends mainly
on its oxydation and spin states, on its coordination number (the
radius increases with the number of neighbors) and on the
polyhedral description of its environment, can be roughly
estimated as a function of these sole criteria and independently
of the detailed structure \cite{Shannon}. It is therefore
interesting to compare the results of the present experiment
concerning the structural distortion of the first shell around the
impurities with the pair distances evaluated from these ionic
radii (tabulated in Table I of Ref. \cite{Shannon}). This
qualitative analysis allows checking the validity of the EXAFS
analysis and lifting some remaining ambiguities.

In practice, the $M$-O pair distance ($M$ denoting the absorbing
ion on the Cu or Ge site) is evaluated by summing the ionic radii
of $M$ and of its O$^{2-}$ first neighbors for comparison with the
distances obtained from EXAFS. The results are summarized in Table
III. The simplest case concerns the Ge site in which Ge$^{4+}$ and
Si$^{4+}$ ions are surrounded by a slightly distorted and a
regular tetrahedron, respectively. The calculations of the Ge-O
and Si-O distances determined from EXAFS are in excellent
agreement with the distances derived from the ionic radii
\cite{Shannon}.

\begin{table*}
\caption{Distances to the first O neighbors, in \AA, calculated
from the ionic radii ($^a$, from Ref. \cite{Shannon}) in
octahedral and in square planar environments (when possible) for
the Cu site, and in tetrahedral environment for the Ge site. The
distances in bold are compatible with those derived from EXAFS and
reported below ($^b$) with the corresponding elongation ratio of
the octahedron. The 2\% and 6\% Ni-doping results are reported.
The last line of both tables gives the superexchange $\alpha$
angle in degrees, estimated from these EXAFS results (see Sec.
\ref{IV-B}).} \label{table3} \vspace{0.5 cm}
\begin{tabular}{l|ccccc|cc}
\hline\hline  &Cu&Ni&Mg&Zn&Mn&Ge&Si \\ \hline planar
square$^a$&{\bf1.92}&1.84&&&&&\\
octahedron$^a$&2.08&{\bf2.04}&{\bf2.07}&{\bf2.09}&{\bf2.18}&&\\
tetrahedron$^a$&&&&&&{\bf1.74}& {\bf1.61}\\ \hline
O(2)$^b$&1.942&1.96/1.98&&2.03&2.08&1.737&1.61\\
O(1)$^b$&2.78&2.26/2.33&&2.36&2.38&1.778&1.61\\${r(M/Cu-O(1))
\over r(M/Cu-O(2))}$ &1.43&1.15/1.18&&1.16&1.14&&\\ \hline
$\alpha$ angle & 98.9(1)&97.9(3)&&
95.8(3)&95.85(15)&98.9(1)&94(1)\\ \hline\hline
\end{tabular}
\end{table*}

The case of the Cu site is more complex. The Cu$^{2+}$ is
surrounded by an elongated octahedron made of a distorted basal
square of 4 O(2) and 2 apical O(1) at a much larger distance. The
calculated Cu-O distance from the ionic radii, in the hypothesis
of square planar configuration, yields a very good agreement with
the measured Cu-O(2) distance indicating that the orbital
configuration of the Cu is mainly sensitive to the 4 closer O(2).
According to the EXAFS results, there are two indistinguishable
solutions for the geometry of the oxygen octahedra around
Ni$^{2+}$ and Zn$^{2+}$ impurities on the Cu site. The apical O(1)
would get closer to the impurity or further away as compared to
the Cu environment, corresponding to an environment closer to a
regular octahedron or closer to a square planar environment,
respectively. Note however that a square planar configuration
exists only for Cu and Ni. Concerning the Ni, the only way to
reproduce the observed increase of the Ni-O(2) distance, as
compared to the Cu one, is an orbital configuration compatible
with an O octahedron around the Ni and excludes a square planar
description. This conclusion also holds for Zn, for which only an
octahedral environment is expected to exist. For Mn$^{2+}$, the
EXAFS analysis was unambiguous. A Mn-O distance close to the one
derived from the EXAFS analysis is indeed found when the ionic
radius of Mn$^{2+}$ is evaluated assuming an octahedral
environment and a high spin state ({\it i.e.} S=5/2), as inferred
from the fine structure splitting of the ESR spectra at low
temperature \cite{grenier3}. Note that the $M$-O distances for Ni,
Zn and Mn, are slightly smaller than the one derived from the
ionic radii because the octahedron is distorted. Some
characteristic features of elongated octahedral environments are
also observed in the X-ray Absorption Near Edge Structure spectra
measured at the Cu K-edge in the pure compound and at the Cu-site
impurities K-edge in the doped compounds, in agreement with the
present analysis. These considerations allow to validate the EXAFS
solutions for the $M$-O distances and to reproduce qualitatively
their increase when substituting Cu by Ni, Zn and Mn, in this
order. For the non-magnetic Mg impurity, extensively studied but
for which no EXAFS measurements could be done, the following
prediction can be made based on the same ionic radii analysis: the
distortions around Mg are expected to be intermediate between
those induced by Ni and Zn. Note that the distortion reduction of
the oxygen octahedron, when replacing Cu by the other impurities,
could be expected since the axial elongation in the Cu$^{2+}$ case
is related to a strong Jahn-Teller effect, usual for this ion. Its
substitution with non-magnetic or non Jahn-Teller 3d ions, like
Ni$^{2+}$ or high spin Mn$^{2+}$, should reduce this effect. The
remaining octahedral distortion is then only induced by the
structural environment.

To summarize, i- the Cu environment is closer to a square planar
one ($r_{Cu-O(1)}/r_{Cu-O(2)} \simeq$ 1.4) while substituting this
site yields an environment closer to a regular octahedron
($r_{M-O(1)}/r_{M-O(2)} \simeq$1.14-1.18 for $M$ = Ni, Zn, Mn)
with the first distance increasing with respect to Cu in the
following order: $M$ = Ni, Mg, Zn, Mn, ii- Si-doping on the Ge
site induces a contraction of its surrounding oxygen tetrahedron,
as strong as the pushing away of the O(2) first neighbors around
the Mn impurity.

\subsection{Superexchange angle calculation}
\label{IV-B}

The distance variations in the impurities environment determined
from the present experiment can now be used to calculate a
structural parameter relevant for the magnetic properties: the
superexchange angle $\alpha$, connecting 2 adjacent Cu sites
through the O(2) atom. In the case of doping on the Cu site, this
angle $\alpha = M-$O(2)$-$Cu will be modified within one chain on
both sides of the impurity $M$, while in the case of doping on the
Ge site, this angle $\alpha =$ Cu$-$O(2)$-$Cu involves the O(2)
connected to the Si impurity and will thus be altered on two
adjacent chains along the $b$ direction (see Fig.\ \ref{Struc2}).

\begin{figure}
\[\includegraphics[scale=0.7]{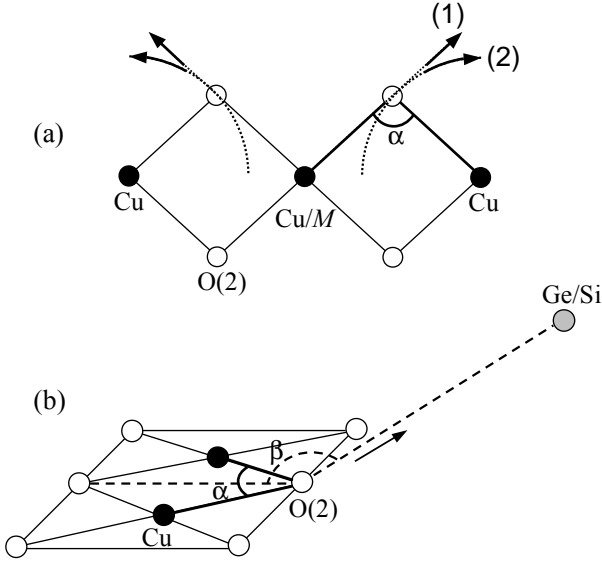}\]
\caption{Schematic view of the local deformation, due to $M$/Cu
(a) and Si/Ge (b) substitution, assumed for the calculation of the
local Cu(or $M$)-O(2)-Cu angle $\alpha$ (see text for the
different scenarii).} \label{struc4}
\end{figure}

Concerning the Cu-site impurities $M=$ Zn, Ni and Mn, two
hypothesis for the direction of displacement of their nearest
neighbors can be made in order to estimate $\alpha$ from the
measured $r_{M-O(2)}$ and $r_{M-Cu}$ distances. Note that the
distance $r_{M-Cu}$ is altered only in the case of Mn doping. In
the first hypothesis, the O(2) are pushed away from the impurity
in the same direction than the initial $Cu$-O(2) one (cf. Fig.\
\ref{struc4}(a), arrow (1)). This also produces a slight
modification of the distance between the O(2) and the first Cu
neighbor of the impurity. In the second hypothesis, this Cu-O(2)
distance is held identical to that in the pure sample, provided a
small modification of the $M$-O(2) direction is made (cf. Fig.\
\ref{struc4}(a), arrow (2)). This assumption is suggested by the
rigidity of the Cu-O(2) bond whose length is found unchanged in
the otherwise largely distorted CuSiO$_3$ compound
\cite{otto,baenitz}. The calculated $\alpha$ values with both
hypothesis are 98-97.7$^{\circ}$ for 2\% Ni, 97.6-97.2$^{\circ}$
for 6\% Ni, 96.1-95.5$^{\circ}$ for 4\% Zn, and 95.7-96$^{\circ}$
for 1.5\% Mn respectively, yielding the average values listed in
Table III. The angle decreases for all impurities on the Cu site,
with respect to their value in the pure compound (99$^{\circ}$),
but this effect is smaller in Ni-doped compounds than in the Zn
and Mn-doped compounds.

In the case of substitution by Si on the Ge site, a complication
arises from the fact that the whole geometry is not planar {\it
i.e.} the $\beta$ angle between the Cu-O(2) ribbons and the
Ge-O(2) segment is equal to 159.5$^{\circ}$ (cf. Fig.\
\ref{struc4}(b)). A simple assumption for the Si-induced
distortion is a contraction of the O tetrahedron around the
impurity without change of bond direction with respect to the Ge
environment. In this case, the O(2) atom is pulled towards the Si
atom by 0.13 \AA\ (arrow in Fig.\ \ref{struc4}(b)). This leads to
an increase of the $\beta$ angle (161.7$^{\circ}$) and a decrease
of the $\alpha$ angle down to 94.1$^{\circ}$. Note that a
different variation of $\beta$ towards larger (resp. smaller)
values would increase (resp. decrease) the value of $\alpha$
(estimated variation within $\pm$ 1$^{\circ}$). A possible
contraction of the Cu-Cu distance was not taken into account in
this simple estimation of $\alpha$, while it is suggested from the
study of Si-doped polycrystals by X-ray diffraction (variation of
the $c$ parameter) \cite{weiden}. This would decrease even more
the Cu-O(2)-Cu angle. In conclusion, the reduction of the $\alpha$
angle should be even stronger for Si than for Zn and Mn-doping.

\section{Discussion}

\subsection{State of the art}

The next step is to relate the EXAFS results, and especially the
above determination of the superexchange angle, to the magnetic
properties of doped CuGeO$_3$. Let us recall that there is a
universal decrease of $T_{SP}$ versus the doping concentration
with $y\approx 3x$ in (Cu$_{1-y}M_y$)(Ge$_{1-x}$Si$_x$)O$_3$,
whereas the lower temperature onset of an AF LRO occurs at
different $T_N$ values depending on the impurity \cite{grenier1}.
Two observations have stimulated a lot of theoretical work
\cite{fukuyama,melin,khomskii2,khomskii3,poilblanc,laflorencie}:
the coexistence of both AF and SP LRO for small Si, Zn, Ni and Mn
doping, as shown by neutron scattering studies
\cite{LPR,Sasago,Coad,grenier4}, and the apparent absence of
critical concentration of impurities for the onset of this AF LRO
\cite{manabe}. From the different models that have been proposed,
a generally admitted description of the magnetic excitation of
these dimerized chains emerges in terms of topological defects
connecting regions of opposite dimerization parity. These defects,
that can be introduced through doping, are then constituted by a
soliton, pinned to the Cu substituent, followed on the same chain
by an antisoliton which restores the interchain phasing. Although
mobile in the isolated chains, the antisoliton remains confined
close to the impurity site when interchain couplings are accounted
for in order to minimize the size of the defective segment which
presents an out-of-phase dimerization with respect to the 3D
pattern \cite{kojima,khomskii2,poilblanc,laflorencie}. Such a
description of the solitonic defects, in terms of
soliton-antisoliton pairs carrying an uncompensated spin 1/2,
leads to large AF fluctuations coexisting with underlying reduced
dimerization. In pure samples, the spin-Peierls phase is
established when the critical size and phase coherence of the
dimerized domains is reached, allowing the system to collapse into
a long-range ordered state. This average size of coherent
dimerized domains is limited by thermally activated and/or
doping-induced solitonic defects. This qualitatively explains the
decrease of T$_{SP}$ with increasing impurity concentration. On
the other hand, in the presence of inter-chain coupling, the onset
of an AF LRO occurs when the AF correlated regions associated with
these doping-induced soliton-antisoliton pairs overlap.

The magnetic behavior under doping is thus qualitatively explained
for non-magnetic impurities within the Cu chains. They simply cut
one dimer, thus releasing one spin 1/2 solitonic defect confined
close to the impurity. However, this is less obvious in the case
of magnetic impurities on the Cu site, which may still be
magnetically coupled to the neighbor Cu spins, and even less for
Si impurities that are introduced outside the Cu chains. The
nature of the released soliton-antisoliton pair through doping is
a key feature in the understanding of the establishment of the AF
LRO. Its nature will crucially depend on the strength of the local
magnetic interaction. It is therefore necessary to relate the
structural parameters determined in the vicinity of the impurity
in the previous sections with the value of the Cu-$M$ exchange
interaction. The variation of the in-chain nearest neighbor
constant, $J_1$, with the Cu-O(2)-Cu superexchange angle $\alpha$
was inferred from the pressure dependence of the susceptibility in
CuGeO$_3$ and found to be equal to: $\delta
J_1/(J_1\delta\alpha)\approx$5.8\% \cite{buchner,khomskii4}.
Various calculations based on the analysis of microscopic
magnetoelastic coupling \cite{werner} and microscopic structural
models \cite{braden} yielded different estimations: $\delta
J_1/(J_1\delta\alpha)\approx$ 10\% and 22\%, respectively. A value
of $\delta J_1/(J_1\delta\alpha)$ ranging between 10 and 20\% was
also found in microscopic calculations of the exchange constants
carried out using different approaches: perturbation theory, exact
diagonalization of Cu$_2$O$_2$ clusters, and band calculations
\cite{khomskii4}. The general result of these studies is that
$\alpha$ reaches its critical value, corresponding to the
cancellation of $J_1$, {\it i.e.} to the transition from AF to FM
coupling, between 90 and 95$^{\circ}$. Other structural parameters
can also affect $J_1$ like the angle $\beta$ between the Cu-O(2)
ribbons and the Ge-O(2) segment, or the pair distances involved in
the exchange path, and more generally, all the parameters that may
influence the strength of the orbital hybridization.

\subsection{Antiferromagnetism vs distortions}

\subsubsection{In-chain magnetic impurities}

Concerning the magnetic impurities on the Cu site, let us recall
that they drive the system towards an AF LRO at a temperature
$T_N$ different from that induced by non-magnetic Cu-site
impurities, slightly lower for Ni and larger for Mn. From our
estimation of the local $\alpha$ angle (Table III), the magnetic
local coupling $J_1^{loc}$ between Ni$^{2+}$ (spin 1) and its
Cu$^{2+}$ neighbors in Ni-doped CuGeO$_3$ is expected to be still
AF but smaller than the coupling between Cu neighbors
($J_1=120-180 K$). Both an analysis of the thermal dependence of
the magnetic susceptibility \cite{grenier1} and a resonance model
for ESR results \cite{grenier2} have indeed evidenced the
formation of Ni-Cu pairs with an AF coupling $J_1^{loc}$ weaker
than $J_1$ (most probable value of $J_1^{loc}$ estimated to be
around $75$ K \cite{grenier2}). The spin 1 of the Ni impurity and
the released spin 1/2 of the Cu therefore form an effective spin
1/2 solitonic defect at low temperature. In the case of Mn-doping,
the $\alpha$ value for the Mn-O(2)-Cu path (Table III) suggests
that $J_1^{loc}$, although still AF, becomes very weak, which
could be reinforced by the frustrating $J_2$ next nearest neighbor
interaction. Here again, some ESR measurements \cite{grenier3}
support this result with the observation of an hyperfine spectrum
of the $^{55}$Mn (nuclear spin I=5/2) with fully developed 6 lines
for very low Mn concentration (0.1 \%). An upper value of
$J_1^{loc} \simeq$4 mK can be estimated from the splitting of the
hyperfine lines. The detection of quasi free 5/2 spins between
$T_{SP}$ and $T_N$ thus demonstrates that the effective coupling
between Mn$^{2+}$ and Cu$^{2+}$ is close to zero. Hence, the onset
of the AF LRO is mainly due to a collective behaviour of the spins
1/2 solitonic defects, decoupled from the free 5/2 spins of the Mn
impurities. It is indeed excluded for the direct Mn exchange to be
responsible for the transition to AF LRO, due to the very large
distances between the Mn atoms.

Therefore, the AF LRO transition observed for all the in-chain
impurities presently studied seems to be mainly due to the
cooperative behavior of soliton-antisoliton pairs (one released
per impurity) carrying a spin 1/2, whatever the magnetic state of
the impurity. Now, another parameter can be considered in order to
explain the differences between the various $T_N(y)$ transition
lines. The present study has revealed differences in the amplitude
of the distortion around each kind of impurities related to their
ionic radii. For instance, the first neighbor distances to the
impurity increases with respect to the Cu environment following
the sequence Ni/Mg/Zn/Mn (section \ref{IV-A}). The strongest local
distortion, in the case of Mn, clearly affects further neighbors
(more than two Cu-Cu distances). It is interesting to note that
the $T_N$ values somehow vary in the same way. The distortions
induced by the impurities along the Cu chains can be very large
(Table II) compared to the atomic displacements produced by the
dimerization itself : 0.2\% relative variation of the first Cu-Cu
distance. In consequence, they will destroy locally the in-chain
dimerization and will render the energetic criterion for
inter-chain phase coherence ineffective. This could in turn affect
the spatial distribution of the soliton-antisoliton pairs along
the chain axis, {\it i.e.} their overall size or/and the mobility
of the antisoliton. It would become easier for the antisoliton to
accommodate its distance from the impurity and the global size of
the soliton-antisoliton pair could be enhanced. As a result, this
additional degree of freedom may favor the overlap of the AF
correlated regions and thus increase the $T_N$. This mechanism
provides an explanation for the reported higher $T_N$ value for Mn
with respect to Zn due to the larger structural distortion of its
environment. On the contrary, for Ni, the distortions are the
smallest implying the lowest $T_N$, as observed. Furthermore, the
AF coupling between the spin 1 Ni and the spin 1/2 Cu may pin the
global spin 1/2 soliton-antisoliton pair closer to the impurity.
As a result, the overlap between the AF correlations is reduced
and the $T_N$ is lowered. Lastly, close enough to T$_N$, one can
not rule out an indirect polarisation of the AF correlated regions
by the Mn$^{2+}$ spin (via the inter-chain interactions for
instance), which could accelerate the transition towards the AF
LRO.

\subsubsection{Out-of-chain Si impurity}

An unsolved issue remains concerning the strong influence of the
out-of-chain Si-doping on the magnetic properties of CuGeO$_3$.
The present study has allowed us to determine the large local
decrease of the superexchange angle $\alpha$ due to Ge
substitution by the much smaller Si ion. $\alpha$ was found to be
smaller than the one calculated in the case of Mn-doping for which
an almost vanishing $J_1^{loc}$ was inferred (Table III). This
strongly suggests that $J_1^{loc}$, coupling the spins of each Cu
pair on both chains facing the Si impurity, is indeed
ferromagnetic. This is in agreement with theoretical predictions
of a weakly ferromagnetic value of $J_1^{loc}$, which was
calculated using realistic structural parameters:
$\alpha$=95$^{\circ}$ and Si-O bond length smaller by 0.13
\AA~than the Ge-O one \cite{khomskii1,khomskii4}. The authors have
invoked additional parameters enhancing this trend towards
ferromagnetism: the further weakening of the hybridization of the
2$p_y$ O(2) orbitals with Si, with respect to Ge, and the
influence of frustrating $J_2$ next nearest interaction. This
result is also comforted by the structure and magnetic properties
of an isostructural CuSiO$_3$ single crystal \cite{otto,baenitz}.
The $\alpha$ angle is equal to 94-95$^{\circ}$ in this compound
\cite{otto,wolfram}, {\it i.e.} close to the one determined
locally around Si impurities in doped CuGeO$_3$ (cf. section
IV.B). The analysis, from magnetometry measurements and a neutron
powder diffraction experiment, of the peculiar AF LRO arising
around 8 K in CuSiO$_3$, suggests a weak ferromagnetic nearest
neighbor exchange interaction $J_1$ and a stronger
antiferromagnetic next nearest neighbor interaction $J_2$, similar
to the CuGeO$_3$ one \cite{otto,wolfram}.

Important consequences are expected from a ferromagnetic
$J_1^{loc}$ on the nature of the solitonic defects produced by the
Si impurity and thus on the associated magnetic properties. The
dimerization pattern imposes that the Si impurity faces one strong
bond corresponding to a dimer and one weak bond between two
dimers. One can reasonably assume that this local variation of
$J_1$ will alter only the strong bond, freeing two spins 1/2, and
will have no effect on the other chain (even locally reinforcing
its dimerization). In this case, the two released spins 1/2 of the
dimer, ferromagnetically coupled, may combine into an effective
free spin 1.

The Curie constant, extracted from the temperature dependence of
the susceptibility between $T_{SP}$ and $T_N$, in a series of
CuGe$_{1-x}$Si$_x$O$_3$ samples \cite{grenierSi}, was previously
analyzed according to Eq. \ref{equation3} in a simple model where
local spins $S$=1/2 are associated with the solitonic defects.
\begin{equation}
\chi_{para}(T)=K_{para}~\frac{(g~\mu_B)^2}{3k_B(T-\theta)}~S(S+1)
\label{equation3}
\end{equation}
\noindent with $g=2.05$, the Land\'e factor, $\mu_B$ the Bohr
magneton, $k_B$ the Boltzman constant, $\theta$ the Curie-Weiss
temperature, and $K_{para}=N~x$ with $N$ the number of released
effective spins $S$ per impurity. Once the contribution of spins
1/2 due to intrinsic impurities has been subtracted, the number
$N$ of spins 1/2 per Si is found close to 3, a puzzling value
\footnote{Note that in Ref. \cite{grenierSi}, the values of
$K_{para}$ were not corrected for the 0.11\% intrinsic spin 1/2
impurities found in pure CuGeO$_3$ and expected to be also present
in doped samples.}. Assuming now that the solitonic defects are
ferromagnetic pairs of spin 1/2, the susceptibility of one pair is
given by :
\begin{equation}
\chi_{pair}(T)=\frac{(g~\mu_B)^2}{k_B
T}\frac{2}{3+\exp(-J_1^{loc}/k_{B}T)} \label{equation4}
\end{equation}
At low temperature, the expression of $\chi_{para}(T)$ then
reduces to Eq. \ref{equation3} for $N$ such magnetic entities with
an effective spin $S$=1. This analysis yields $N\sim~$1.2, which
is close enough to the expected value of one ferromagnetic pair of
spins 1/2 per Si impurity in this new description. This can be
further argued from the analysis of the magnetization measured in
a 0.3\% Si-doped CuGeO$_3$ sample (see Fig. 2 of Ref.
\cite{grenier5}). The $M(H)$ curve reaches a first saturation
plateau $M_{sat}=Nxg\mu_BS=0.0077 \mu_B$ in the SP phase before
entering the incommensurate phase. The analysis of the whole
$M(H)$ curve with Brillouin functions, after correction for the
intrinsic impurities contribution, yields $N\sim~$1.1 spins 1 or
$N\sim~$2.2 spins 1/2. The first result is in better agreement
with the expected number of effective spins 1 and with the
previous estimation of $N$, deduced from susceptibility analysis.
The revised analysis of the magnetic measurements in Si-doped
compounds is therefore in favor of the release of one solitonic
defect per Si with an effective spin 1, on one of both adjacent
chains.

A consequence of the effective spin 1 of the solitonic defect
produced by Si will be to renormalize $T_N\propto S(S+1)$ by a
factor of 2.67 as compared to the case of spin 1/2 solitonic
defects released by the same concentration of Cu-site impurities.
However, this comparison only holds if the spatial extension of
the solitonic defect is otherwise identical, thus producing the
same overlap between AF correlated regions. The overall size of
the soliton-antisoliton pair along the $c-$axis was indeed
estimated to be the same for Si and Zn from X-ray measurements
\cite{pouget2}. This supported the origin of the factor $\sim$3
between the $T_N$ of Si and of non-magnetic Cu-site impurities
from the different types of the effective spin associated to the
solitonic defects, at least for low doping levels (see Fig.\
\ref{Diag}). The reason why this scaling does not hold at higher
$x$ may be due to a different mechanism involved in the onset of
AF LRO. The distances between impurities are becoming shorter than
the 1D magnetic correlations, and moreover, a first-order
transition is observed in the case of Cu-site impurities between a
dimerized and a uniform AF LRO, whereas the coherence length of
the spin-Peierls ordering decreases progressively in the Si doping
case \cite{grenier6}.

\subsection{Spin-Peierls transition vs distortions}

Contrary to the sensitivity of the AF LRO to the nature of the
impurity, discussed in the previous section on the basis of the
EXAFS results, the decrease of the spin-Peierls temperature
appears to be more universal. On the Cu site, the decrease of
$T_{SP}$ has exactly the same concentration dependence for all
studied Cu substituents, whatever the spin impurity and the
amplitude of the in-chain local distortions around impurities (see
Fig.\ \ref{Diag}). For Si-doping on the Ge site, $T_{SP}$
decreases $\sim~$3 times faster with the concentration, although
the size of the solitonic defect induced by Si in the chain
direction is similar to the one induced by Zn impurities
\cite{pouget2}.

Moreover, the average correlation length of the dimerized regions
along $c$ does not seem to be so relevant for the appearance of
the SP LRO, as shown by the measurements of fluctuations in pure
and doped CuGeO$_3$ for different impurities above the SP
transition \cite{schoeffel,pouget2}. First, the regime of 1D
pretransitional fluctuations along the chain direction in pure
CuGeO$_3$ was shown to start well above $T_{SP}$, while these
fluctuations become 2D and then 3D much closer to $T_{SP}$.
Secondly, the average size of the dimerized regions along $c$ in
the pretransitional regime for Si, is comparable to that of Zn
despite the factor $\sim~$3 in their $T_{SP}$. Last, the extension
of the 1D pretransional fluctuations along $c$ is much smaller for
Ni than for Zn or Mg \cite{pouget2}, whereas their decrease of
$T_{SP}$ has exactly the same doping level dependence.

Therefore, all these experimental results point out that the
relevant parameter for the decrease of $T_{SP}$ through doping is
not the way the dimerization is affected along the chain
direction, but perpendicular to it through the stabilization of
the out-of-phase 3D dimerization pattern. This is supported by a
major difference observed concerning the spatial extension of the
perturbation of the dimerized regions when doping inside and
outside of the Cu chains: Only one chain is affected for in-chain
doping by Zn instead of three chains in the case of out-of-chain
Si-doping \cite{pouget2}. This has provided Pouget {\it et al.}
\cite{pouget2} with a simple model explaining the factor 3,
between Si and Zn doping, in the respective decrease of $T_{SP}$
and in the critical concentration above which the SP LRO is
destroyed. According to the present study, we can generalize this
last result concerning Zn to the other Cu substituents, Ni and Mg,
which produce smaller local distortions. As concerns the Mn
impurity, which leads to stronger distortions than Zn, the fact
that $T_{SP}$ has the same concentration dependence tends to show
that Mn-doping affects only one Cu chain. Note that the furthest
analyzed atomic distance probed by our EXAFS experiment is shorter
than that to the next Cu on a neighboring chain (4.25~\AA). In
summary, the universality of the SP transition seems to be related
to the transverse weakening of the dimerization which depends
therefore essentially on the position of the impurity, outside or
inside the chains.

Lastly, let us come back to the difference in the distortions
induced by different concentrations of the same impurity. The
Ni-O(2) distance to the first neighbors increases more for 6\% Ni
than for 2\% Ni doping with respect to the Cu environment. The
higher doped compound, also, has a more disordered environment. It
is interesting to note that these concentrations lie before and
after the end of the $T_{SP}$ line, and on each side of the
transition from dimerized to uniform AF LRO. This result
underlines some influence of the disorder in the disappearance of
the spin-Peierls phase.

\section{Conclusion}

In the present study, the EXAFS technique was used as a powerful
probe of the local atomic arrangement around various impurities in
the paramagnetic state of doped CuGeO$_3$. This technique has
allowed us to quantify the distortions in the impurity
environment, such as the modification of the shape of the O
octahedron around the Cu site, and to evaluate the local
modification of the angles and distances relevant for the magnetic
properties. An increase of the first neighbor distance to the Cu
site was observed when substituting Cu by Ni, Zn and Mn, in this
order. An elongation of the distances to the further neighbors was
evidenced in the case of Mn-doping which thus produces the largest
alteration among the Cu site impurities. The angle $\alpha$ of the
Cu-O-Cu superexchange path leading to the AF $J_1$ interaction
between Cu$^{2+}$ first neighbors was found to decrease
progressively when doping with Ni, then with Zn or Mn, and at last
with Si. These results have allowed establishing the local
weakening of the $J_1$ interaction for Ni/Cu substitution, its
almost cancellation in the case of Mn/Cu substitution and, lastly,
its change of sign for Si doping on the Ge site, leading to a
small ferromagnetic coupling between each set of Cu pairs of two
adjacent chains.

In a global scheme where each impurity is considered to release
magnetic solitonic defects, an analysis of the structural
distortions of the doped CuGeO$_3$ in relation with their magnetic
properties could then be proposed. A good description of the
magnetic behavior of Si-doped compounds is obtained calling upon
ferromagnetic pairs of spins 1/2, which yields in particular a
simple explanation for the renormalization of $T_N$ for Si with
respect to Zn-doping. Another interesting consequence of this
study concerns the Cu-site impurities for which the amplitude of
the structural distortion, rather than the magnetic moment of the
impurity, is found to be correlated to the variation of $T_N$.
This last result underlines the importance of the structural
distortions in the magnetic behavior of spin-Peierls doped
compounds and opens the way to models taking into account
realistic structural parameters.

We would like to thank B. Canals, P. Monod, R. M\'elin, D.
Khomskii and N. Laflorencie for fruitful discussions.

\end{document}